# Ultrafast 1 MHz vacuum-ultraviolet source via highly cascaded harmonic generation in negative-curvature hollow-core fibers

David E. Couch[1,*], Daniel D. Hickstein[2,5], David G. Winters[2], Sterling J. Backus[2,3], Matthew S. Kirchner[2], Scott R. Domingue[2], Jessica J. Ramirez[2], Charles G. Durfee[4], Margaret M. Murnane[1], and Henry C. Kapteyn[1,2]

[1]*University of Colorado at Boulder, Department of Physics and JILA, Boulder, CO 80309, USA*
[2]*Kapteyn-Murnane Laboratories Inc., 4775 Walnut St #102, Boulder, CO 80301, USA*
[3]*Colorado State University, Ft. Collins, CO 80523, USA*
[4]*Colorado School of Mines, Department of Physics, Golden, CO 80401, USA*
[5]*dhickstein@kmlabs.com*
*\*David.Couch@colorado.edu*

**Abstract:** Vacuum ultraviolet (VUV) light is critical for the study of molecules and materials, but the generation of femtosecond pulses in the VUV region at high repetition rates has proven difficult. Here, we demonstrate the efficient generation of VUV light at MHz repetition rates using highly cascaded four-wave mixing processes in a negative-curvature hollow-core fiber. Both even and odd order harmonics are generated up to the 15$^{th}$ harmonic (69 nm, 18.0 eV), with high energy resolution of ~40 meV. In contrast to direct high harmonic generation, this highly cascaded harmonic generation process requires lower peak intensity and therefore can operate at higher repetition rates, driven by a robust ~10 W fiber-laser system in a compact setup. Additionally, we present numerical simulations that explore the fundamental capabilities and spatiotemporal dynamics of highly cascaded harmonic generation. This VUV source can enhance the capabilities of spectroscopies of molecular and quantum materials, such as photoionization mass spectrometry and time-, angle-, and spin-resolved photoemission.

## 1.   Introduction

The vacuum ultraviolet (VUV) spectral region, covering approximately 6–15 eV, has a unique ability to probe physical and chemical processes. For example, the bond energies of all molecules and the ionization energies of most materials lie in this energy range. Consequently, VUV light sources are used as an ionization source in angle-resolved photoelectron spectroscopy (ARPES) [1–5] and photoionization mass spectrometry (PIMS) [6–12], or to initiate chemical reactions relevant to atmospheric science in a controlled environment [13–15]. However, this science is limited by the lack of bright VUV light sources, especially coherent (laser) sources. While low-flux sources, such as deuterium lamps, can satisfy some applications [16], many experiments require higher flux to overcome shot noise in a reasonable amount of time. Single wavelength sources, such as the 9$^{th}$ harmonic of a Nd:YAG laser (118.2 nm, 10.49 eV) [17], are not tunable, limiting the scope of experiments for which they can be used. Tunable deep ultraviolet light has been produced from phase-matched four-wave mixing schemes [18,19], but extending this strategy into the VUV has typically provided very limited or no tunability [20–22]. Synchrotrons and free electron lasers are currently the only sources of fully tunable, high-flux VUV light [8]; however, these facility-scale sources have limited access and time resolution.

Direct high-harmonic generation (HHG) driven by intense femtosecond laser pulses can generate multiple harmonic orders throughout the VUV, extreme UV (EUV), and soft X-ray (SXR) spectral regions when very high intensities (>10$^{13}$ W/cm$^2$) are available. However, scaling to higher repetition rates with lower pulse energy is not as simple as focusing tighter to reach the same peak intensities. In a free-focus geometry, small focal spot sizes correspond to impractically short lengths and high gas pressures [23], while for HHG in a hollow capillary waveguide, the confinement loss scales with the inverse cube of the core diameter [24]. HHG using solid [25,26] and liquid [27] targets has recently been investigated with good success. However, only a few papers [27,28] have demonstrated generation into the VUV, and the path to scaling the flux and efficiency to enable applications is still unclear.

Negative curvature anti-resonant hollow-core fibers offer an attractive alternative to a simple hollow capillary waveguide, as they use microstructures in the core region to induce interference effects that confine light to a small diameter with minimal propagation loss [29–31]. Notably, HHG has been demonstrated in a similar anti-resonant photonic crystal fiber at 1 kHz repetition rate [32]. Nevertheless, using MHz-repetition-rate fiber lasers, it is still impractical to achieve peak intensities high enough for efficient conversion using direct HHG, without rapid damage to the core microstructures.

Here we utilize a new highly-cascaded harmonic generation (HCHG) process to enable the simultaneous production of thirteen UV/VUV spectral lines using driving laser intensities well below the threshold required for HHG [33,34]. By focusing two colors (the fundamental and second harmonic of a 10 W average power Yb:fiber laser) into a xenon-filled negative curvature hollow-core fiber, and tuning the xenon pressure to provide optimal phase-matching, both even and odd harmonic orders are generated ranging from the 3$^{rd}$ to the 15$^{th}$ harmonic order. We use a peak intensity of approximately $2 \times 10^{12}$ W/cm$^2$, which is significantly lower than the ~$1 \times 10^{14}$ W/cm$^2$ typically required for efficient HHG [35]. Indeed, at the intensity used in our experiments, the ponderomotive energy of the electron is only 0.1 eV and the single-atom, single-color ionization probability [36,37] is less than $10^{-12}$ – far below the typical HHG regime. As expected for a perturbative cascaded interaction, the flux at each harmonic decreases with increasing harmonic order, but at ~$10^{14}$–$10^{17}$ ph/s it rivals synchrotron flux levels [8] for photon energies up to 10.8 eV. Using a model that only includes the third order nonlinearity ($\chi^{(3)}$), we confirm our hypothesis regarding the cascaded mechanism for harmonic generation and gain a deeper understanding of the physics of HCHG.

## 2. Experiment

We start with a prototype ultrafast ytterbium-doped fiber laser producing 10-µJ, 160-fs pulses at 1035 nm with a repetition rate of 1 MHz. The output is then frequency doubled to 518 nm (BBO, Eksma Optics) with about 50% efficiency. The 1035- and 518-nm pulses are overlapped in time using a delay stage and focused into a 30-µm diameter negative-curvature fiber (Glo-Photonics PMC-500/700) filled with ~1000 Torr (20 psia) of xenon gas (Fig. 1a). With 4 W of 1035-nm light and 3 W of 518-nm light entering the fiber (parallel linear polarization), we generate about 200 mW of third harmonic (345 nm) through a degenerate four-wave mixing process – here two photons of the second harmonic are combined to produce one photon each of the fundamental and the third harmonic. This third harmonic light can then combine with the fundamental and second harmonic light to drive additional four-wave-mixing processes in a cascaded series (Fig. 1b). Such cascaded processes have been studied previously, but past studies [38–40] have used Ti:Sapphire lasers with shorter pulses, order-of-magnitude higher pulse energies, and larger diameter capillary waveguides. In contrast, we employ a compact fiber laser and a small diameter negative-curvature waveguide to produce harmonics up to the 15$^{th}$ harmonic.

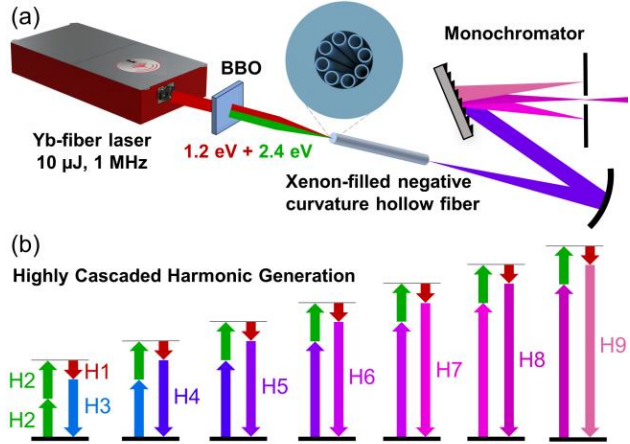

**Fig. 1. Highly cascaded harmonic generation (HCHG).** a) In our experiment, a 1035-nm laser is frequency doubled and both fundamental and second harmonic beams are focused into a xenon-filled, 30-µm core diameter, hollow-core negative-curvature fiber to drive the HCHG process. The resulting VUV light is then spectrally selected by a grating-based monochromator. b) The HCHG process is the result of numerous four-wave mixing steps, each combining three photons to generate a higher-energy photon. We note that the photon combination for H4 and higher represent just one possible route to each harmonic.

After generation, the VUV light passes through an in-vacuum monochromator in order to select a single wavelength for transmission to the sample. In these experiments, we employed three different monochromator designs. (1) First, we used a prism monochromator to measure the flux of harmonic orders 3-9. In this setup, the VUV beam was collimated using at concave collimating mirror (R=500 mm), diffracted through a MgF$_2$ prism (60° apex), reflected by a flat wavelength-tuning mirror, focused (R=400 mm) through a 1 mm exit slit, before reaching one of several detectors. All reflective optics were coated with MgF$_2$-protected aluminum optimized for 120 nm (Acton

Optics and Coatings). The power of harmonic orders 3–5 were measured on a thermal power meter (Newport), while the power of the 9th harmonic was measured on a calibrated $Al_2O_3$ photodiode (NIST). The relative power of orders 5–9 was measured by replacing the detector with a Ce:YAG crystal (1 mm thick) imaged onto a camera (Edmund Optics). The power of orders 6–8 was estimated by calibrating the relative power measurement (camera) to the absolute power measurement (photodiode) of 9th harmonic. Using this method, the estimated flux of the 5th harmonic agreed to within a factor of two with the power meter measurement, suggesting that our flux estimates for orders 6–8 are also accurate to within a factor of two. (2) For measuring a high-resolution spectrum of the 8th and 9th harmonics, we replaced the prism and rotating flat mirror with a reflective diffraction grating (Richardson Grating Lab, 1200 g/mm, $MgF_2$ coating) which could be rotated to tune wavelength. We used a 50 μm exit slit and the $Al_2O_3$ detector. This 90° Czerny-Turner monochromator has an estimated resolution of 10 meV. (3) Finally, for harmonic orders 9–15, we eliminated the focusing optic after the diffraction grating in the Czerny-Turner design, instead using a single (R=250 mm, Acton 120-nm coating) optic to refocus the beam onto the 1 mm exit slit, with a converging beam incident on the grating (Wadsworth monochromator, Fig. 1). A bare aluminum grating (Richardson Grating Lab, 1200 g/mm) was used to increase reflectivity at wavelengths below the $MgF_2$ absorption edge near 110 nm. We also included a rejector mirror (angle of incidence 72°) in the beam, placed immediately after the exit of the hollow fiber. This dielectric mirror was designed to transmit most of the 1035-, 517-, and 345-nm light while reflecting all shorter wavelengths. This mirror greatly reduces the heat load on the more sensitive Al+$MgF_2$ coated optics. This also allowed us to direct the 1035-, 517-, and 345-nm beams outside the vacuum chamber to monitor for optimal fiber coupling and temporal overlap.

### 3. Results and Discussion

Using highly cascaded harmonic generation, we simultaneously generate bright even and odd harmonic orders 3–15 of the bi-chromatic 1035- and 517-nm driving laser. The measured radiant flux (radiant power) and the corresponding photon flux of each harmonic are recorded in Table 1 and shown in Fig. 2. High resolution spectra of the 8th and 9th harmonics (Fig. 2a, shown in red) reveal bandwidths of 40 meV (full width at half maximum), which is somewhat more bandwidth than the 8-meV bandwidth of the driving laser, indicating that the HCHG process likely produces shorter pulses, or pulses that can be compressed to be shorter, than the driving pulses. This type of temporal shortening is often seen with nonlinear optical processes such as HHG [41].

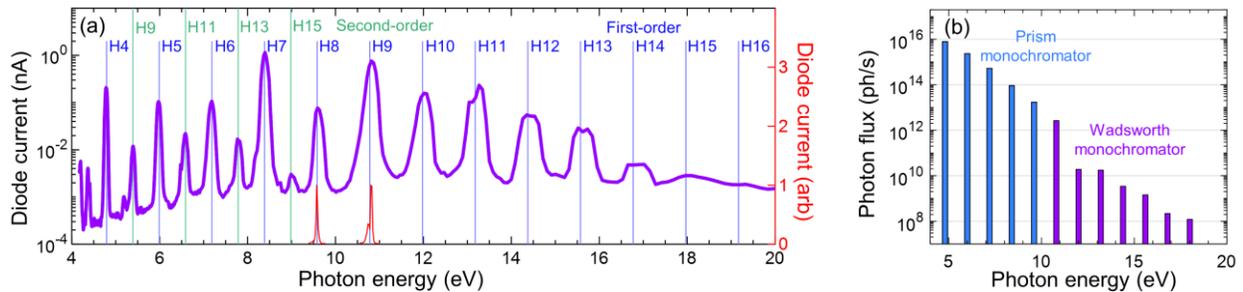

**Fig. 2. Spectrum and calibrated photon flux observed for each harmonic.** a) VUV spectra acquired using the high-throughput (purple, log scale) and high-resolution (red, linear scale) monochromator configurations. Harmonics up to H14 are clearly visible. The high-resolution monochromator reveals an intrinsic spectral resolution of 40 meV ($E/\Delta E_{FWHM} \approx 250$ for H8 and H9). b) Observed photon flux for each harmonic, measured using the prism monochromator (4–10 eV, blue) and the Wadsworth monochromator (10–18 eV, purple). Above 11 eV, the beamline optics have substantially reduced efficiency.

**Table 1. Observed and estimated source photon flux for each harmonic.** Harmonic orders 3–9 were measured using a prism-based monochromator, while harmonics 9–15 were measured using a grating-based Wadsworth monochromator. The estimated photon flux at source for harmonics 10–15 should be regarded as an order-of-magnitude estimate and is limited by knowledge of the precise reflectivity of the optics.

| Harmonic order | Photon energy (eV) | Wavelength (nm) | Power (μW) | Measured photon flux (ph/s) | Estimated flux from source (ph/s) |
|---|---|---|---|---|---|
| 3 | 3.6 | 345 | 230000 | $4.0 \times 10^{17}$ | $4.7 \times 10^{17}$ |
| 4 | 4.8 | 259 | 6000 | $7.8 \times 10^{15}$ | $1.6 \times 10^{16}$ |
| 5 | 6.0 | 207 | 2200 | $2.3 \times 10^{15}$ | $4.6 \times 10^{16}$ |
| 6 | 7.2 | 173 | 610 | $5.3 \times 10^{14}$ | $1.6 \times 10^{15}$ |
| 7 | 8.4 | 148 | 120 | $9.1 \times 10^{13}$ | $2.7 \times 10^{14}$ |
| 8 | 9.6 | 129 | 26 | $1.7 \times 10^{13}$ | $6.8 \times 10^{13}$ |
| 9 | 10.8 | 115 | 4.6 | $2.6 \times 10^{12}$ | $2.1 \times 10^{13}$ |
| 10 | 12.0 | 104 | 0.036 | $1.9 \times 10^{10}$ | $3 \times 10^{12}$ |
| 11 | 13.2 | 94 | 0.037 | $1.7 \times 10^{10}$ | $3 \times 10^{12}$ |
| 12 | 14.4 | 86 | 0.0079 | $3.4 \times 10^{09}$ | $9 \times 10^{11}$ |
| 13 | 15.6 | 80 | 0.0036 | $1.4 \times 10^{09}$ | $3 \times 10^{11}$ |
| 14 | 16.8 | 74 | 0.00059 | $2.2 \times 10^{08}$ | $5 \times 10^{10}$ |
| 15 | 18.0 | 69 | 0.00035 | $1.2 \times 10^{08}$ | $3 \times 10^{10}$ |

The HCHG process demonstrated here is a cascaded four-wave-mixing process. This is distinct from HHG, where each harmonic is generated directly from the driving laser using high-order nonlinearities. The perturbative approach here uses the lowest order isotropic nonlinear polarizability term of xenon, $\chi^{(3)}$, to generate all observed harmonics. The first step of our harmonic generation process is the generation of 3$^{rd}$ harmonic from the 2$^{nd}$ harmonic and fundamental beams (Fig. 1b). The energy conservation is described by:

$$\omega_3 = \omega_2 + \omega_2 - \omega_1, \quad (1)$$

where $\omega_n$ is the angular frequency of the $n$-th harmonic, $\omega_n = n\omega_1$, and $\omega_1$ is the frequency of the driving laser ($\hbar\omega_1 = 1.2$ eV). The phase mismatch in a gas-filled hollow waveguide [42] is given by:

$$\Delta k \equiv k_1 + k_3 - 2k_2 = 2\pi N \left(\frac{\delta_3}{\lambda_3} + \frac{\delta_1}{\lambda_1} - \frac{2\delta_2}{\lambda_2}\right) - \frac{u}{4\pi a^2}(\lambda_3 + \lambda_1 - 2\lambda_2), \quad (2)$$

where $\lambda_n$ is the n$^{th}$ harmonic wavelength, k$_n$ is the corresponding wavevector, N is the number density of the gas, $\delta_n$ is the gas dispersion (related to the refractive index by $n - 1 = N\delta$) for the n$^{th}$ harmonic, u is a mode-dependent constant (2.405 for the lowest order mode used here [42]), and $a$ is the diameter of the waveguide. The xenon refractive index has been measured [43] for wavelengths longer than 140 nm. The first term on the right side of Eq. 2 is the pressure-dependent contribution from the gas, and the second term is the pressure-independent term from the waveguide confinement. Typically, there is some pressure that will eliminate the phase mismatch, allowing the 3$^{rd}$ harmonic generation to be phase matched.

Once the 3$^{rd}$ harmonic is produced, this wavelength can lead to the generation of higher harmonics by the same process. Beyond the 4$^{th}$ harmonic, multiple pathways can lead to the production of each harmonic, for example:

$$\begin{aligned}
\omega_4 &= \omega_3 + \omega_2 - \omega_1 \\
\omega_4 &= \omega_3 + \omega_3 - \omega_2 \\
\omega_5 &= \omega_4 + \omega_2 - \omega_1 \\
\omega_5 &= \omega_3 + \omega_3 - \omega_1 \\
\omega_6 &= \omega_5 + \omega_2 - \omega_1 \\
\omega_6 &= \omega_4 + \omega_3 - \omega_1 \\
\omega_7 &= \omega_6 + \omega_2 - \omega_1 \\
\omega_7 &= \omega_5 + \omega_3 - \omega_1 \\
\omega_7 &= \omega_4 + \omega_4 - \omega_1 \\
&\ldots
\end{aligned} \quad (3)$$

Each of these processes has an optimal phase-matching condition, and most of these processes are phase-matched for some xenon pressure below that needed to phase-match 3$^{rd}$ harmonic generation, which is the first step in the cascaded process. Because we apply pressure only to the front of the 100-mm fiber (around 1000 Torr, tuned to optimize harmonic production) and allow the gas to flow through the fiber into vacuum, we have a gradient of pressure along the length of the fiber (Fig. 3a,b), estimated from a finite element calculation employing a pressure-dependent conductance. The range of xenon pressures in our fiber allows many of the relevant phase-matching and quasi-phase-matching conditions to be met at some point along the fiber. However, more complicated situations are possible – past work [38,39] indicates that quasi-phase-matching conditions arising from the periodic build-up and decay of intermediate harmonics could be more important than true phase-matching for cascaded processes

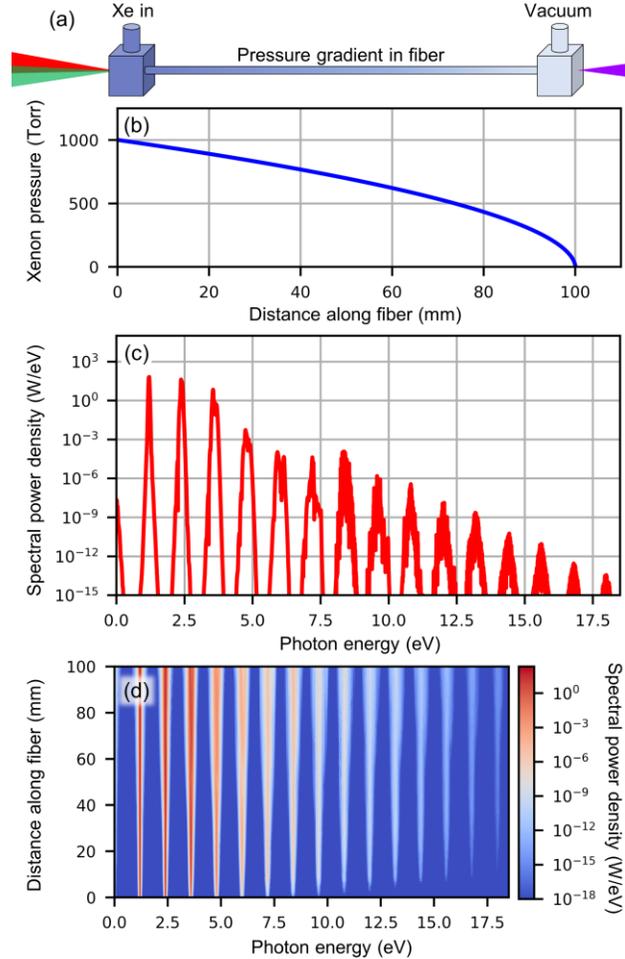

**Fig. 3. Simulation of HCHG.** a) The xenon is supplied to the front of the fiber so that the pressure in the fiber decreases to vacuum along the length of the fiber. b) A finite element simulation of the pressure in the fiber shows that the pressure decreases most rapidly at the end of the fiber, as the flow transitions from viscous flow to molecular flow. c) Using the pressure profile from (b), numerical simulations using the nonlinear Schrödinger equation (NLSE) confirm the generation of numerous harmonics in the UV and VUV spectral region. d) Each harmonic is generated throughout the length of the fiber, rather than within a small phase-matched region.

To provide a more complete picture of the HCHG process, we perform numerical calculations using the nonlinear Schrödinger equation (NLSE) and implemented in the PyNLO package [44,45]. In these simulations, the pulses are modelled with a sech$^2$ temporal profile with a full-width-at-half-maximum of 160 fs and a pulse energy of 2 µJ each. The pressure-dependent nonlinear index of xenon is assumed to be $1.1 \times 10^{-22}$ m$^2$/(W bar) [46,47]. The calculations predict output fluxes (Fig. 3c) roughly comparable to what we measure (Fig. 2), confirming that our experimental results are consistent with a $\chi^{(3)}$-driven cascaded mixing process. These calculations reveal that the harmonics are produced all along the length of the fiber (Fig. 3d), indicating that quasi-phase-matching is likely important in the production of bright harmonics. The pressure profile used here was chosen primarily for experimental convenience and may not be ideal for harmonic generation. Further experimentation and more advanced calculations are needed to determine the ideal pressure profile for HCHG.

One advantage of direct HHG is the ability to generate wavelengths extending far beyond the ionization energy of the nonlinear medium where absorption is very high. In contrast, the cascaded harmonic generation shown here could reasonably be expected to cut off at the first harmonic that is strongly absorbed by the medium. However, an additional calculation shown in Fig. 4 reveals that attenuation of harmonics beyond 3ω have only a small effect on the flux of higher harmonics. Moreover, the experimental results shown in Fig. 2 also indicate that absorption of a higher harmonic does not prevent the generation of higher harmonics. The ionization energy of xenon is 12.13 eV, with a high density of Rydberg states in the vicinity of our 10$^{th}$ harmonic at 12.0 eV. As shown in Fig. 2, we observe a lower intensity at this wavelength, yet harmonics up to the 15$^{th}$ are observed. We therefore conclude that HCHG process has some ability to generate harmonics beyond absorption bands in the medium. Furthermore, in HCHG, we can effectively drive the NLO process with relatively long-duration pulses, giving us the opportunity to generate harmonics with spectral bandwidth significantly narrower than is typical for HHG.

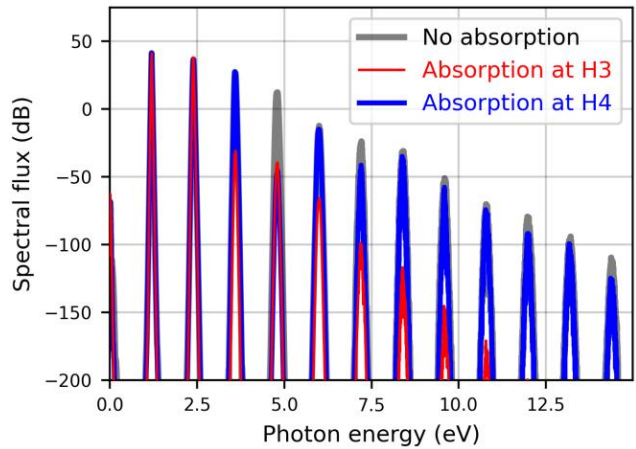

**Fig. 4. Simulated harmonic flux when one harmonic is severely attenuated.** Attenuation of the 3$^{rd}$ harmonic (red) leads to all higher harmonics being severely attenuated. In contrast, attenuation of the 4$^{th}$ harmonic (blue) has only a small effect on the higher harmonic fluxes. This result indicates that multiple pathways can produce each harmonic above the 4$^{th}$ order.

The HCHG process is a powerful tool for bringing high-flux VUV light to tabletop experiments. To our knowledge, this is the smallest, most energy-efficient source of femtosecond pulses of vacuum-ultraviolet light in the 7–11 eV spectral range. The total energy consumption of all electronics and power supplies used for this source is approximately 1 kW (an additional ~1 kW is used by water chillers). Nevertheless, our flux for our harmonics in the 6–10 eV spectral range meets or exceeds that of the Advanced Light Source synchrotron ($2 \times 10^{13}$ photons/s for similar bandwidth). [48] Thus, bright multispectral VUV light can be implemented in individual labs, and synchrotron facilities can then be used more efficiently for experiments requiring fully tunable VUV light or high fluxes at higher photon energies.

The VUV light produced by this source is linearly polarized, which is a valuable feature that broadens the scope of possible applications [49]. Any linear polarization for the VUV can be produced, as the polarization of the driving lasers is conserved. Additionally, the use of circularly polarized driving lasers may allow circularly polarized light to be generated directly. Circularly polarized harmonics have been produced from both HHG (counter-rotating fields) [50] and four-wave mixing (seed co-rotating with a single pump) [51]. Since HCHG is simply a cascade of multiple four-wave mixing steps, it should also produce circularly polarized light when co-rotating fields are used. Future studies will be required to experimentally verify this capability.

This source could enable major advancements in important scientific applications such as time-, spin- and/or angle-resolved photoemission spectroscopy [1–5,52–54] as well as PIMS [6–8,12]. For example, the high repetition rate of this laser is well suited for ARPES experiments that can suffer from a loss of energy resolution due to space charge effects. Moreover, the tunability of this VUV source is currently being used for PIMS experiments, where it has demonstrated the ability to differentiate between different molecules that have the same mass [55].

## 4. Conclusion

We have demonstrated the generation of high-flux vacuum-ultraviolet light using highly cascaded harmonic generation. Numerical simulations show that it is driven by cascaded four-wave mixing processes. The observed spectral bandwidths of ~40 meV are ideal for many scientific applications, and narrower spectra down to 1 meV can be obtained using moderately sized monochromators. The process of highly cascaded harmonic generation can bring bright, high repetition rate, multispectral VUV light, previously available only at large facilities, to laboratory tabletops.

**Funding**

The authors acknowledge funding from the following: Department of Energy (DOE) Basic Energy Sciences (BES) (DE-FG02-99ER14982); Air Force Office of Scientific Research (FA9550-16-1-0121); National Science Foundation (NSF) Graduate Research Fellowship Program (DGE-1650115); DARPA PULSE (W31P4Q-13-1-0015).

**Acknowledgements**

Current affiliations: DEC: Sandia National Laboratories. SJB, MSK, SRD, JJR: Thorlabs Inc. DGW: Lockheed Martin.

**Disclosures**

DDH, SJB, DGW, MSK, SRD, JJR, HCK: KMLabs (E,P).
MMM, HCK, SJB: KMLabs (I).
KMLabs uses this HCHG technology to build VUV laser systems.